\documentclass[12pt,a4paper]{article}
\usepackage[dvips]{graphicx}

\begin{document}
\sloppypar
\bigskip

\title{ Dependence of Nuclear Level Density on Vibrational State Damping}

\author{ V.A. Plujko$^{1,2}$, A.N. Gorbachenko$^1$}

\date{}

\maketitle

\vspace{-1cm}
\begin{center}
$^{1}$Taras Shevchenko National  University,
Glushkova Str. 6, Kiev, Ukraine \\
$^{2}$Institute for Nuclear Research,
Prosp. Nauki 47, Kiev, Ukraine\\
E-mail: plujko@univ.kiev.ua
\end{center}
\vspace{0.1cm}

\begin{center}
\parbox{0.8\textwidth}{\small
The response function approach is proposed to
include vibrational state in calculation of level
density. The calculations show rather strong dependence of
level density on the relaxation times of collective
state damping.}
\end{center}

\section{ Introduction}

Level density $\rho $ is one of the main quantities
to define characteristics of  nuclear decay. The collective states
can strongly effect on level density, specifically, at low excitation
energies \cite{Ignat83}-\cite{Ignat93}.
Calculation of the variation factor $K$ of level densities
is the simplest method to estimate effect of the vibrational
states on level densities.
The factor $K$ is the ratio of level densities with and without allowing
for vibrational states. The level densities can be  calculated
within framework of statistical approach by
standard saddle-point method\cite{Ignat83} or modificated one\cite{Gros}.
It was found\cite{Pluj02} that  factor $K$ is practically independent
of type of statistical approach. Therefore we will use below
standard saddle-point method as a simplest
way for calculation of the  variation factor of level density.

In this contribution work the effect of the vibration
state damping on nuclear level density is studied on the
base of response function approach in wide range of
excitation energies. The dependence of variation factor
$K$ on damping of vibrational state is
investigated.

\section{ Response function method for calculation
of the level density variation}

Variation factor of level density is defined as
\begin{equation}
K={\rho}/{{\rho}_{0}},
\end{equation}
where $\rho$ and ${\rho}_{0}$ are level densities with and
without allowing for vibrational states.

The level density has the following form within framework of
statistical approach in standard saddle-point method\cite{Ignat83}
\begin{equation}
\label{eq1}
\rho \left( {U,A} \right) = \left( {4\pi ^{2}D} \right)^{ - 1/2}
\exp S\left({\alpha _{0} ,\beta _{0}}  \right),
\end{equation}
where
$
S\left( {\alpha _{0} ,\beta _{0}}  \right) =
- \alpha _{0} A + \beta _{0} E
+ \ln Z\left( {\alpha _{0} ,\beta _{0}}  \right)$
is entropy of nucleus with mass number $A$ at excitation energy
$U=E-E_{g.s}$  with $E_{g.s}$ for ground state energy;
$
Z\left( {\alpha ,\beta}  \right)={\rm Tr}
\left[ \exp\left( -\beta \hat{H}+\alpha\hat{A} \right)
\right]
$
- partition function; $\hat {H}$ - nuclear Hamiltonian; $\hat {A}$ -
operator of particle number; $D$ - determinant of second partial
derivatives of logarithm of partition function
with respect to parameters $\alpha$ and $\beta$.
The saddle-points $\alpha _{0} ,\,\,\beta _{0} $ define
temperature ($T = 1/\beta _{0}$) and chemical potential ($\mu =
\alpha _{0} /\beta _{0}$). They are solutions of the system of
thermodynamic state equations
$A={{\partial{\,\ln Z}}/{\partial{\alpha}}}$,
 $E=-{{\partial{\,\ln Z}}/{\partial{\beta}}}$.

We use conception of random phase approximation and
suppose that collective vibrational states of the multipolarity $L$
are formed by the separable residual interaction
$V_{res}^{k}=k \, \hat Q_{L} \hat Q_{L}^{+}$, where
$\hat Q_{L}$ is one-body operator of multipole momentum with
multipolarity $L$, $\hat Q_{L}\propto r^{L} Y_{L\,M} $,
and $k$ is a coupling constant.

The partition function can be presented in the form\cite{Bloch}
$
Z=Z_{0}\cdot\Delta Z ,
$
where $Z_{0}$ is partition function of the independent particle model.
The quantity $\Delta Z$ can be expressed by variation of
thermodynamic potential $\Delta \Omega$ in the form
$\Delta Z = \exp (-\beta \Delta\Omega )$ with
the following equation for $\Delta \Omega$
\begin{equation}
\label{eqr71}
 \Delta \Omega =
\int\limits_{0}^{k} {d{k}'} \left\langle {{\hat Q_{L}
\hat Q_{L}^{+}}} \right\rangle/2
 =\frac{2L+1}{2\pi} \int\limits_{0}^{k} {d{k}'}
\int\limits_{-\infty}^{+ \infty} { \frac{-1} {e^{\beta\varepsilon} - 1}
 Im {\rm X}^{k'}_{L} d\varepsilon }.
\end{equation}
Here, ${\rm X}^{k'}_{L}$ is nuclear response function\cite{DiToro}
on external field proportional to $\hat Q_{L}$ with
frequency $\omega=\varepsilon/\hbar$;
symbol $<...>$ denotes averaging over total Hamiltonian with
residual interaction $V_{res}^{k'}$.
The final expression in right-hand side of the eq.(\ref{eqr71})
is obtained after transformation of relationship for $\Delta\Omega$
with the use of Green function method\cite{Bogol}.
Finally variation of the partition function is presented in the
following form
\begin{equation}
\label{eq5}
\Delta Z\left( {\beta}  \right) \equiv \Delta Z_{vib} = \mathop { \prod}
\limits_{j}
 \left| {\left[ {\frac{{1 - \exp\left( { - \left( {
{E} _{j} ^{0} + i\Gamma _{j} ^{0}}  \right)\beta}
\right)}}{{1 - \exp\left( { - \left( { {E} _{j}  + i\Gamma
_{j} }  \right)\beta}  \right)}}} \right]^{2L + 1}} \right|\,
\end{equation}
\noindent with the use of transformation similar to that one from
Ref.\cite{Ignat74}
Here, $ {E} _{j}  $, $\Gamma _{j}  $ are real
and imaginary parts of  solutions of the equation
\begin{equation}
\label{root1}
k-1/{{\rm X}^{0}_{L}}(\varepsilon)=0,
\end{equation}
and $ {E} _{j}^{0}$, $\Gamma _{j}^{0} $ are real and imaginary
 parts of roots of the equation
$
1/{{\rm X}^{0}_{L}}(\varepsilon)~=~0$ with ${\rm X}^{0}_{L}$
for response function of independent particle
model ($k=0$).

%\vspace{-0.4cm}

\section{  Calculations and conclusion}

Semiclassical Vlasov-Landau approach
for response function\cite{DiToro} was used in the calculations.
The nuclear mean field and relaxation processes
can be easy taken into account in this approach.
Collision integral was considered in relaxation time approach.
 Three different parametrizations of
relaxation time $\tau_{c}$ were adopted with taken into account retardation
effects\cite{Kolomi,PlujAct01}.

1) Expression  within framework of transport theory\cite{Kolomi,PlujAct01}
\begin{equation}
\label{eq10}
\hbar /\tau _{c} \left( {\varepsilon ,T} \right) = \left(
{\left( {\varepsilon/ {2\pi}} \right)^{2} + T^{2}}
 \right)/4.07 \ \ {\rm MeV} .
\end{equation}
\noindent

2)Relaxation time due to model of doorway state
\begin{equation}
\label{eq11}
\hbar /\tau _{c} \left( {\varepsilon ,T} \right) = 2\pi
\left( {\frac{{K_{m}} }{{A^{3}}}}
\right)\frac{{g^{3}}}{{2}}\frac{{\left( {\varepsilon + U } \right)^{2}}}
{{3 + \bar {n}}} \, ,
\end{equation}
\noindent
where $\bar {n} = 0.843\,a\,T$;
$K_{m}$ is defined from agreement between relaxation times given
by eqs.(\ref{eq10}) and (\ref{eq11}) at $T = 0$ and
at GQR energy; $g = g\left( {\varepsilon _{F}}\right)=6a/\pi^2 $ with
level density parameter $a=A/8\,\,$MeV$^{ - 1}$.

3) Model of doorway state
\begin{equation}
\label{eq12}
\hbar /\tau _{c} \left( {\varepsilon ,T} \right) =2\pi
\left({\frac{{K_{B}} }{{A^{3}}}}
\right)\frac{{g^{3}}}{{8}}{{\left( {\varepsilon+ U } \right)}} \, ,
\end{equation}

\noindent
where $K_{B} $ is defined from agreement with relaxation times
given by eqs.(\ref{eq10}) and (\ref{eq12}) at $T = 0$ and
at GQR energy.

The dependence of variation  factor $K$ on excitation energy
is shown on fig.1 for nucleus $^{56}$Fe.  Low-lying
quadrupole state with energy $E _{2^{+},\, exp}
 = 0.847\,\,$MeV was taken into account; histogram is
microscopic calculation and experimental
values\cite{Ignat88}. Solid line shows calculation
with relaxation time given by eq.(\ref{eq10}),
dash - with relaxation time according to eq.(\ref{eq11}),
dotted - eq.(\ref{eq12}) for $\tau_{c}$.

Parametrization of relaxation time given by
eq. (\ref{eq12}) leads to small value of variation factor in
comparison with other parametrization of relaxation times.
Level density rather strongly depends on a relaxation times.
It can give an additional possibility for investigation of the
relaxation time dependence  on temperature and collective state energy.

\vspace{-0,5cm}

{\setlength{\baselineskip}{0.7\baselineskip}

}

\newpage
\pagestyle{empty}

\begin{figure}
\begin{center}
\hspace{-10mm}
 \includegraphics[width=0.8\textwidth,clip]{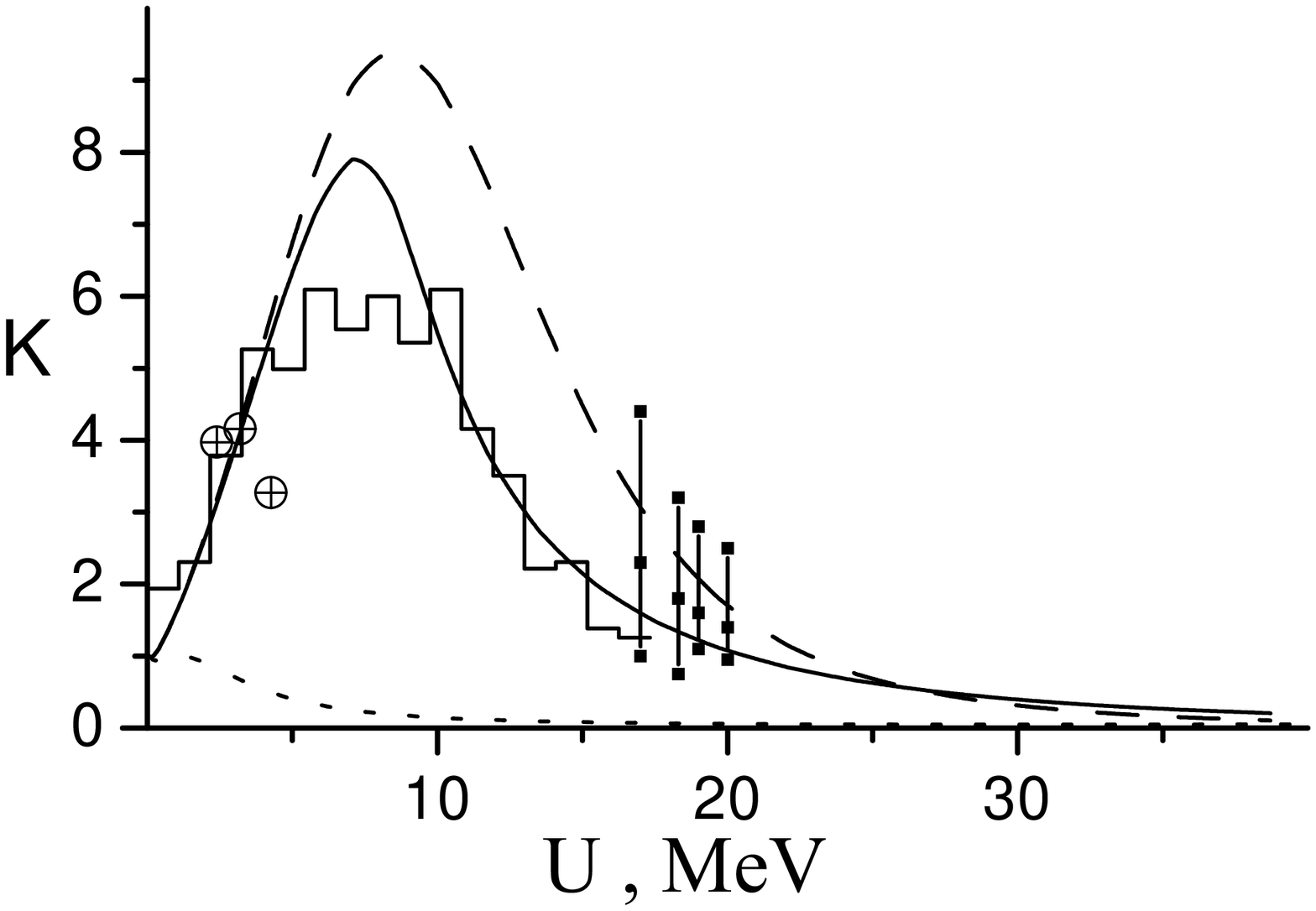}
\vspace{10mm}
 \parbox[t]{0.9\textwidth}{ Fig. 1: Variation factor
of level density for ${}^{56}$Fe at different relaxation times; lines:
solid - relaxation time due to (\ref{eq10}),
dash - (\ref{eq11}), dotted - (\ref{eq12});
histogram - microscopic calculation and experimental
values{\protect\cite{Ignat88}}.}
 \end{center}
\end{figure}

\end{document}